\documentclass[12pt]{iopart}

\begin{document}

\title[Plane wave tunnelling in a square barrier]{Tunnelling of plane waves through a
square barrier}

\author{J. Julve}

\address{IMAFF, Consejo Superior de Investigaciones
Cient\'\i ficas, Serrano 113 bis, Madrid 28006, Spain}

\ead{julve@imaff.cfmac.csic.es}

\author{F. J. de Urr\'{\i}es}

\address{Departamento de F\'\i sica,
Universidad de Alcal\'a de Henares, Alcal\'a de Henares (Madrid),
Spain}

\ead{fernando.urries@uah.es}

\begin{abstract}
The time evolution of plane waves in the presence of a 1-dimensional
square quantum barrier is considered. Comparison is made between the
cases of an infinite and a cut-off (shutter) initial plane wave. The
difference is relevant when the results are applied to the analysis
of the tunnelling regime. This work is focused on the analytical
calculation of the time-evolved solution and highlights the
contribution of the resonant (Gamow) states.



\end{abstract}

\maketitle

\section{Introduction}

The study of the traversal of potential barriers by wave packets in
the 1-dimensional case is relevant for the electron transport
through barrier junctions, the physics of wave guides and light
transmission through Photonic Band Gaps, etc. On more theoretical
grounds, the resonances, transient excitations and the time of
arrival of wave packets, with ensuing paradoxes of non-locality and
super-luminal tunnelling, are further interesting aspects.

Relevant resonances are mostly expected in systems of two or more
barriers separated by a gap. Actually they generally occur in any
simple barrier, and the plain square barrier is most tractable,
fully representative for many theoretical purposes and devoid of
bound states or other unessential features.

On the other hand, practical calculations face important limitations
and the effort of pushing the analytical calculation as far as
possible is always rewarding. Significant progress has been
accomplished \cite{Muga1} for a Gaussian packet impinging on a
square barrier, where the Gaussian structure was exploited, and
\cite{Calderon1} for the shutter problem, where the contribution of
the resonant (Gamow) states in a double square barrier was worked
out.

In this work we consider the time evolution, in the presence of a
square barrier of height $V$ in the interval [$0,L$], of an initial
plane wave with support on the whole 1-D space, aiming to work out
the contribution of the resonant states. Our calculation extends the
results of the shutter problem, where the initial plane wave
occupied the space axis at one side of the barrier and was specially
suited to study details of the propagation of the wave front.

The reason for doing this is twofold. First, the simpler problem of
the shutter initial condition $\psi_s(x)= \theta(-x)\rme^{\rmi kx}$
has been assumed in the literature to be meaningful also for the
analysis of the transmission in the tunnelling regime when $k\ll
k_V\equiv\sqrt{2mV}\;$. Unfortunately such a cut-off initial wave
function is highly non-monochromatic as its Fourier transform shows:
$$\tilde\psi_s(p)=\frac{1}{\sqrt{2\pi}}\int_{-\infty}^{\infty} \rmd x
\, \rme^{-\rmi
px}\psi_s(x)=\frac{1}{\rmi}\frac{1}{\sqrt{2\pi}}\;\rm{PP}\frac{1}{k-p}+\sqrt{\frac{\pi}{2}}
\;\delta(k-p)$$ The Principal Part distribution around $k$ is
exceedingly wide so that the amplitude of the momentum $p=k_V$
skimming the barrier is $\propto(k-k_V)^{-1}$, a sizeable value (in
modulus, actually it is $>k_V^{-1}$) even for very small incident
$k\;$, with a slow decrease for $p>k_V\;$. Secondly, by considering
infinite plane waves the way is paved for further uses, since
superpositions of them build up any desired initial wave packet. In
particular, the compact result obtained for the Gaussian wave packet
\cite{Muga1} could be re-derived pinpointing the contribution of
each single Fourier component.

In this work we follow the approach used by Peierls and
Garc\'{\i}a-Calder{\'o}n \cite{Calderon2}, namely
Laplace-transforming the time evolution equation into a second order
linear differential equation (SOLDE) in one variable, expressing its
solutions in terms of the Green function (GF) with resonant boundary
conditions (RBCs) and then undoing the Laplace transform.

The first step is done in Section 2, where a brief discussion is
made of the use of the Green's method for the solution of SOLDEs
when the boundary conditions (BCs) required for the GF and for the
solution are different. The GF with RBCs features a simple structure
of isolated (resonance) poles, in terms of which the ($p$-dependent)
Laplace-transformed wave function is worked out in Section 3. Then
the explicit calculation of the inverse Laplace transform leading to
the sought after $t$-dependent solution is carried out in Section 4.

As a check of this cumbersome analytical calculation, the $t=0$
(Section 5) and the $t=\infty$ (Section 6) limits of the
time-dependent solution $\psi (x,t)$ are calculated: at $t=0$ one
must recover the initial plane wave and at $t=\infty$ a suitable
stationary solution must be reached. Then the conclusions are drawn
in Section 7. Some notations and a number of calculations and
technical details are deferred to the Appendices.

\section{The evolution equations}

We consider the 1-D time-dependent $\rm Schr\ddot{o}dinger$ equation
\begin{equation}
(\rmi\hbar\frac{\partial}{\partial t}-H)\psi (x,t)=0
\end{equation}
where the Hamiltonian $H=p^2/\,2m\;+V(x)$ corresponds to the square
barrier potential $V(x)=\theta(x)\theta(L-x)V$ and the solution
satisfies the initial condition $\psi (x,0)\equiv\psi_o(x)=
\rme^{\rmi kx}\;,\,k>0\, $.

The Laplace transform $\bar{\psi}(s)=\int_o^\infty \rmd t\,\,
\rme^{-st}\psi(t)$ on the time variable, brings the parabolic
partial derivative differential equation (1) to the simpler SOLDE
\begin{equation}
L_x\bar\psi(x,p)\equiv[\frac{\partial^2}{\partial
x^2}+p^2-\frac{2m}{\hbar^2}V(x)]\;\bar\psi(x,p)
=\rmi\alpha\;\rme^{\rmi kx}
\end{equation}
where $\alpha\equiv 2m/{\hbar}$ and $p^2=\rmi\alpha\,s$ (Appendix
A). Besides the usual scattering \emph{in} and \emph{out} solutions,
the time-independent $\rm Schr\ddot{o}dinger$ homogeneous equation
$L_x\,\bar\psi=0$ has the resonant solutions satisfying the
homogeneous outgoing RBCs
\begin{equation}
\partial_x \bar\psi\!\!\mid_{x=0}\,=-\rmi p\,\bar\psi(0) \;\;,\;\;
\partial_x \bar\psi\!\!\mid_{x=L}\,=\rmi p\,\bar\psi(L)\;,
\end{equation}
which exist only for a denumerable set of isolated values $p_n$ of
$p\;$ lying in the lower half complex plane (Appendix A).

A solution to the inhomogeneous equation (2) for $p^2 \in R_{+}$ can
be explicitly written down for each region of $V(x)$ :
\begin{eqnarray}
\bar\psi_{I}(x,p)&=&B\,\rme^{-\rmi px}+\frac{\rmi\alpha}{p^2-k^2}\rme^{\rmi kx}\hskip 2.7cm(x<0)\nonumber\\
\bar\psi_{II}(x,p)&=&M\,\rme^{\rmi p'x}+N\,\rme^{-\rmi p'x}+\frac{\rmi\alpha}{p'^2-k^2}\rme^{\rmi kx}\hskip 0.5cm(0\leq x\leq L)\;\\
\bar\psi_{III}(x,p)&=&A\,\rme^{\rmi
px}+\frac{\rmi\alpha}{p^2-k^2}\rme^{\rmi kx}\hskip 3cm(x>L)\nonumber
\end{eqnarray}
where the amplitudes $A$, $B$, $M$ and $N$ are functions of $p$
completely determined by the matching conditions at $x=0$ and $x=L$
and have a common denominator of the form
$D(p)\equiv\ominus^2\rme^{\rmi p'L}-\oplus^2\rme^{-\rmi p'L}$
(notation in Appendix A). Out of all the terms of the general
homogeneous solution, the choice of $\rme^{\rmi px}$ (for $x>L$) or
$\rme^{-\rmi px}$ (for $x<0$) is dictated by the behaviour of the
solution for $x\rightarrow\pm\infty$ , since a small positive
imaginary part in $p$ is supposed when performing the inverse
Laplace transform leading to $\psi(x,t)\;,\; t>0$ (Appendix D).

However, this expression of the solution is not suited to perform
the inverse Laplace transform back to $\psi(x,t)$ because of the
nontrivial analytical form of the amplitudes above. Instead the GF
approach lets us to express $\bar\psi(x,p)$ as a sum of isolated
pole terms plus other simple terms easier to deal with. To this end,
(2) together with the Green equation for $G(x,x',p)$ can be written
in the form
\begin{eqnarray}
L_x\bar\psi(x,p)-\rmi\alpha \rme^{\rmi kx}&=&0\\
L_xG(x,x',p)-\delta(x-x')&=&0\;,
\end{eqnarray}
where $G(x,x',p)$ is required to obey RBCs, namely
\begin{equation}
\partial_x G(x,x',p)\!\!\mid_{x=0}\,=-\rmi p\,G(0,x',p)\;\;,\;\;
\partial_x G(x,x',p)\!\!\mid_{x=L}\,=\rmi p\,G(L,x',p)
\end{equation}
It is expected that $G(x,x',p)=L^{-1}_x$ will have poles at
$p=p_n\;$, where the homogeneous equation $L_x\bar\psi=0$ with RBCs
has non-trivial solutions.

The equations (5) and (6) may conveniently be given the short-hand
notation $\Psi\!=\!0$ and $\Gamma\!=\!0$ respectively. Then the
Green method uses the integral equation $\int_{0}^{L}\rmd
x\;[\bar\psi\Gamma-G\Psi]=0$ to obtain $\bar\psi(x,p)$ in terms of
$G(x,x',p)$. Only if $\bar\psi$ and $G$ obey the same
\emph{homogeneous} BCs, the surface terms in this integral cancel
out and one obtains the familiar result
$\bar\psi_{II}(x,p)=\int_{0}^{L}\rmd
x'\,\rmi\alpha\,G(x',x,p)\,\rme^{\rmi kx'}$.

However the BCs for $\bar\psi(x,p)$ (involving $\bar\psi$ and
$\partial_x\bar\psi$ both at $x=0$ and at $x=L$) stemming from (4)
are different from (3) and non-homogeneous. In that case one obtains
a modified expression for the region $II$, namely
\begin{eqnarray}
\fl\bar\psi_{II}(x,p)\;\;=&&\int_{0}^{L}\rmd x'\,\rmi\alpha\,G(x',x,p)\,\rme^{\rmi kx'}\nonumber\\
&&-\frac{\alpha}{p+k}G(L,x,p)\rme^{\rmi
kL}-\frac{\alpha}{p-k}G(0,x,p)
\end{eqnarray}
For the external regions $I$ and $III$, the matching of $\bar\psi$
at $x=0$ and $x=L$ determine the coefficients $B$ and $A$
respectively, so that
\begin{eqnarray}
\fl\bar\psi_{I}(x,p)\;\;=&&\rmi\alpha\,\rme^{-\rmi px}\int_{0}^{L}\rmd x'\,\rmi\alpha\,G(x',0,p)\,\rme^{\rmi kx'}\nonumber\\
&&-\frac{\alpha}{p+k}G(L,0,p)\rme^{\rmi kL}\rme^{-\rmi px}-\frac{\alpha}{p-k}G(0,0,p)\rme^{-\rmi px}\\
&&-\frac{\rmi\alpha}{p^2-k^2}\rme^{-\rmi
px}+\frac{\rmi\alpha}{p^2-k^2}\rme^{\rmi kx}\nonumber
\end{eqnarray}
and
\begin{eqnarray}
\fl\bar\psi_{III}(x,p)\;\;=&&\rmi\alpha\,\rme^{\rmi p(x-L)}\int_{0}^{L}\rmd x'\,\rmi\alpha\,G(x',L,p)\,\rme^{\rmi kx'}\nonumber\\
&&-\frac{\alpha}{p+k}G(L,L,p)\rme^{\rmi p(x-L)}\rme^{\rmi kL}-\frac{\alpha}{p-k}G(0,L,p)\rme^{\rmi p(x-L)}\\
&&-\frac{\rmi\alpha}{p^2-k^2}\rme^{\rmi p(x-L)}\rme^{\rmi
kL}+\frac{\rmi\alpha}{p^2-k^2}\rme^{\rmi kx}\nonumber
\end{eqnarray}
For comparison, only the last term in the r.h.s. of (8), the last
three ones in (9) and the third one in (10) appear in the shutter
problem.

The crucial advantage of the method is that almost all the Green
functions involved in the equations above can be expanded as a sum
of terms which are simple poles in $p_n$.

\section{Analytical structure of the p-dependent solution}

General theorems \cite{Nussenzveig} and the explicit analytical
derivation (Appendix B) of $G(x,x',p)$ show that $|G(x,x',p)|\sim
1/|p|\,\rightarrow0$ as $|p|\rightarrow\infty$ in the complex plane
for almost any $x\in[0,L]$ and $x'\in[0,L]$ . Then the
Mittag-Leffler theorem tells that
\begin{equation}
G(x,x',p)=\sum_{n}\frac{C_n(x,x')}{p-p_n}
\end{equation}
The exception is for $G(0,0,p)$ and $G(L,L,p)$, the modulus of which
grows as $|p|$ for $|p|\rightarrow\infty$ in the lower half complex
plane $p$ and requires some special care (Appendix C).

The residues in the r.h.s of (11) can be easily computed
\cite{Calderon2} and one finds $C_n(x,x')=u_n(x)u_n(x')/N_n$ , where
the functions $u_n(x)$ belong to the denumerable set of the resonant
solutions satisfying $L_x\,u_n(x)=0$ with RBCs
$$\partial_xu_n(x) \!\!\mid_{x=0}\,=-\rmi p_n\,u_n(0) \;\;,\;\;
\partial_x u_n(x)\!\!\mid_{x=L}\,=\rmi p_n\,u_n(L)$$
and $N_n$ are suitable normalization factors (Appendix A).

The inverse Laplace transform
\begin{equation}
\fl\psi(x,t)=\frac{1}{2\pi \rmi}\int_{c-\rmi \infty}^{c+\rmi
\infty}\rmd s\;\rme^{st}\bar\psi(x,p(s)) = \frac{1}{2\pi
m}\int_{-\infty}^{+\infty}\rmd
p\;p\;\rme^{-\rmi\frac{p^2}{2m}t}\bar\psi(x,p)\;,
\end{equation}
written as an integral over the real momentum variable $p\,$
(Appendix D), leads one to consider the pole expansion of
$p\;\bar\psi(x,p)$ in the integrand, which can be worked out for
each region $I$, $II$ and $III$ (Appendix E):
\begin{eqnarray}
\fl p\;\bar\psi_{I}(x,p)\;\;=&&\rmi\alpha\,\rme^{-\rmi px}\sum_{n}\frac{p}{p-p_n}\frac{u_n(0)}{N_n}\int_{0}^{L}\rmd x'\,u_n(x')\,\rme^{\rmi kx'}\nonumber\\
&&+\alpha\frac{k}{p+k}G(L,0,-k)\rme^{-\rmi px}\rme^{\rmi kL}-\alpha\sum_{n}\frac{p_n}{p-p_n}\frac{1}{k+p_n}\frac{u_n(L)u_n(0)}{N_n}\rme^{-\rmi px}\rme^{\rmi kL}\nonumber\\
&&+\alpha\frac{p}{k}G(0,0,0)\rme^{-\rmi px}-\alpha\frac{1}{k}\frac{p^2}{p-k}G(0,0,k)\rme^{-\rmi px}\\
&&+\alpha\sum_{n}\frac{p^2}{p-p_n}\frac{1}{p_n}\frac{1}{k-p_n}\frac{u_n^2(0)}{N_n}\rme^{-\rmi
px}+\frac{\rmi\alpha}{2}(\frac{1}{p-k}+\frac{1}{p+k})(\rme^{\rmi
kx}-\rme^{-\rmi px})\nonumber
\end{eqnarray}

\begin{eqnarray}
\fl p\;\bar\psi_{II}(x,p)\;\;=&&\rmi\alpha\sum_{n}\frac{p}{p-p_n}\frac{u_n(x)}{N_n}\int_{0}^{L}\rmd x'\,u_n(x')\,\rme^{\rmi kx'}\nonumber\\
&&+\alpha\frac{k}{p+k}G(L,x,-k)\rme^{\rmi kL}-\alpha\,\rme^{\rmi kL}\sum_{n}\frac{p_n}{p-p_n}\frac{1}{k+p_n}\frac{u_n(L)u_n(x)}{N_n}\\
&&-\alpha\frac{k}{p-k}G(0,x,k)+\alpha\sum_{n}\frac{p_n}{p-p_n}\frac{1}{k-p_n}\frac{u_n(0)u_n(x)}{N_n}\nonumber
\end{eqnarray}

\begin{eqnarray}
\fl p\;\bar\psi_{III}(x,p)\;\;=&&\rmi\alpha\,\rme^{\rmi p(x-L)}\sum_{n}\frac{p}{p-p_n}\frac{u_n(L)}{N_n}\int_{0}^{L}\rmd x'\,u_n(x')\,\rme^{\rmi kx'}\nonumber\\
&&-\alpha\frac{p}{k}G(L,L,0)\rme^{\rmi p(x-L)}\rme^{\rmi kL}+\frac{p^2}{p+k}\frac{\alpha}{k}G(L,L,-k)\rme^{\rmi p(x-L)}\rme^{\rmi kL}\nonumber\\
&&-\alpha\sum_{n}\frac{p^2}{p-p_n}\frac{1}{p_n}\frac{1}{k+p_n}\frac{u_n^2(L)}{N_n}\rme^{\rmi p(x-L)}\rme^{\rmi kL}\\
&&+\alpha\sum_{n}\frac{p_n}{p-p_n}\frac{1}{k-p_n}\frac{u_n(0)u_n(L)}{N_n}\rme^{\rmi p(x-L)}+\alpha\frac{k}{k-p}G(0,L,k)\rme^{\rmi p(x-L)}\nonumber\\
&&+\frac{\rmi\alpha}{2}(\frac{1}{p+k}+\frac{1}{p-k})(\rme^{\rmi
kx}-\rme^{\rmi p(x-L)}\rme^{\rmi kL})\nonumber
\end{eqnarray}
In the third row of (14) and in the fourth row of (15) one
recognizes the terms of the shutter problem given in
\cite{Calderon1}. The shutter terms in the region I, not calculated
in \cite{Calderon1}, are given in the third and fourth rows of (13).

\section{The time-dependent solution}

For each of the terms in the expressions above, the integrals
stemming from (12) can be brought to the form of an integral
representation of the error function $er\!f\!c(z)$ so that their
inverse Laplace transform can be carried out thoroughly (Appendix
F). We obtain:
\begin{eqnarray}
\fl\psi_{I}(x,t)\;\;=&&-\sum_{n}\frac{u_n(0)}{N_n}(\int_{0}^{L}\rmd
x'\,u_n(x')\,\rme^{\rmi kx'})\;[A^{I}_n]\nonumber\\
&&+\rmi k G(L,0,-k)\rme^{\rmi kL}\;[B^{I}_{-k}]-\rmi\rme^{\rmi kL}\sum_{n}\frac{p_n}{k+p_n}\frac{u_n(L)u_n(0)}{N_n}\;[B^{I}_n]\\
&&+\rmi G(0,0,0)\frac{1}{k}\;[S^{I}_0]-\rmi G(0,0,k)\frac{1}{k}\;[S^{I}_k]+\rmi\sum_{n}\frac{1}{p_n}\frac{1}{k-p_n}\frac{u_n^2(0)}{N_n}\;[S^{I}_n]\nonumber\\
&&-\frac{1}{2}\rme^{\rmi kx}\;[S^{I}_1] -\frac{1}{2}\rme^{\rmi
kx}\;[S^{I}_2]+\frac{1}{2}[S^{I}_3]+\frac{1}{2}[S^{I}_4]\nonumber
\end{eqnarray}
where the factors in square brackets embody the time
($\tau=t\,/\,2m$) dependence of the solution and are the result of
the integrations over the momentum $p\,$, namely:
\begin{eqnarray}
A^{I}_n&=&-\rme^{\rmi\frac{x^2}{4\tau}}(p_n\rme^{y^2_{-x,n}}\;er\!f\!c(y_{-x,n})-\frac{\rme^{\rmi\frac{\pi}{4}}}{\sqrt{\pi\tau}})\nonumber\\
B^{I}_{-k}&=&\rme^{\rmi\frac{x^2}{4\tau}}\rme^{y^2_{-x,-k}}\;er\!f\!c(y_{-x,-k})\nonumber\\
B^{I}_n&=&\rme^{\rmi\frac{x^2}{4\tau}}\rme^{y^2_{-x,n}}\;er\!f\!c(y_{-x,n})\nonumber\\
S^{I}_0&=&\frac{\rme^{\rmi\frac{\pi}{4}}}{\sqrt{\pi\tau}}\frac{x}{2\tau}\rme^{\rmi\frac{x^2}{4\tau}}\nonumber\\
S^{I}_k&=&-\rme^{\rmi\frac{x^2}{4\tau}}(k^2\rme^{y^2_{-x,k}}\;er\!f\!c(y_{-x,k})-\frac{\rme^{\rmi\frac{\pi}{4}}}{\sqrt{\pi\tau}}
(k-\frac{x}{2\tau}))\\
S^{I}_n&=&-\rme^{\rmi\frac{x^2}{4\tau}}(p_n^2\rme^{y^2_{-x,n}}\;er\!f\!c(y_{-x,n})-\frac{\rme^{\rmi\frac{\pi}{4}}}{\sqrt{\pi\tau}}(p_n-\frac{x}{2\tau}))\nonumber\\
S^{I}_1&=&-\rme^{-\rmi\tau{k^2}}\;er\!f\!c(\rmi\sqrt{\rmi\tau}k)\nonumber\\
S^{I}_2&=&-\rme^{-\rmi\tau{k^2}}\;er\!f\!c(-\rmi\sqrt{\rmi\tau}k)\nonumber\\
S^{I}_3&=&-\rme^{\rmi\frac{x^2}{4\tau}}\rme^{y^2_{-x,k}}\;er\!f\!c(y_{-x,k})\nonumber\\
S^{I}_4&=&-\rme^{\rmi\frac{x^2}{4\tau}}\rme^{y^2_{-x,-k}}\;er\!f\!c(y_{-x,-k})\nonumber
\end{eqnarray}
where we have introduced the variables
$y_{x,q}\equiv\rme^{-\rmi\frac{\pi}{4}}(4\tau)^{-\frac{1}{2}}(x-2\tau
q)$ for $q=p_n,k,-k$.

Likewise
\begin{eqnarray}
\fl\psi_{II}(x,t)\;\;=&&-\sum_{n}\frac{u_n(x)}{N_n}(\int_{0}^{L}\rmd x'\,u_n(x')\,\rme^{\rmi kx'})\;[A^{II}_n]\nonumber\\
&&+\rmi kG(L,x,-k)\rme^{\rmi kL}\;[B^{II}_{-k}]
-\rmi\rme^{\rmi kL}\sum_{n}\frac{p_n}{k+p_n}\frac{u_n(L)u_n(x)}{N_n}\;[B^{II}_n]\\
&&-\rmi kG(0,x,k)\;[S^{II}_k] +\rmi
\sum_{n}\frac{p_n}{k-p_n}\frac{u_n(0)u_n(x)}{N_n}\;[S^{II}_n]\nonumber
\end{eqnarray}
where
\begin{eqnarray}
A^{II}_n&=&-p_n\,\rme^{-\rmi\tau{p_n^2}}\nonumber\\
B^{II}_{-k}&=&-\rme^{-\rmi\tau{k^2}}\;er\!f\!c(-\rmi\sqrt{\rmi\tau}k)\nonumber\\
B^{II}_n&=&-\rme^{-\rmi\tau{p_n^2}}\;er\!f\!c(\rmi\sqrt{\rmi\tau}p_n)\\
S^{II}_k&=&-\rme^{-\rmi\tau{k^2}}\;er\!f\!c(\rmi\sqrt{\rmi\tau}k)\nonumber\\
S^{II}_n&=&-\rme^{-\rmi\tau{p_n^2}}\;er\!f\!c(\rmi\sqrt{\rmi\tau}p_n)\nonumber
\end{eqnarray}
Finally
\begin{eqnarray}
\fl\psi_{III}(x,t)\;\;=&&-\sum_{n}\frac{u_n(L)}{N_n}(\int_{0}^{L}\rmd x'\,u_n(x')\,\rme^{\rmi kx'})\;[A^{III}_n]\nonumber\\
&&-\rmi G(L,L,0)\frac{\rme^{\rmi kL}}{k}\;[B^{III}_0]
+\rmi G(L,L,-k)\frac{\rme^{\rmi kL}}{k}\;[B^{III}_{-k}]\nonumber\\
&&-\rmi\rme^{\rmi kL}\sum_{n}\frac{1}{p_n}\frac{1}{k+p_n}\frac{u_n^2(L)}{N_n}\;[B^{III}_n]\\
&&-\rmi kG(0,L,k)\;[S^{III}]+\rmi
\sum_{n}\frac{p_n}{k-p_n}\frac{u_n(0)u_n(L)}{N_n}\;[S^{III}_n]
\hskip 3cm\nonumber\\
&&-\frac{1}{2}\rme^{\rmi kx}\;[C^{III}_1] -\frac{1}{2}\rme^{\rmi
kx}\;[C^{III}_2] +\frac{1}{2}\rme^{\rmi
kL}\;[C^{III}_3]+\frac{1}{2}\rme^{\rmi kL}\;[C^{III}_4]\nonumber
\end{eqnarray}
where
\begin{eqnarray}
A^{III}_n&=&-\rme^{\rmi\frac{(x-L)^2}{4\tau}}(p_n\rme^{y_{x-L,n}^2}\;er\!f\!c(y_{x-L,n})+\frac{\rme^{\rmi\frac{\pi}{4}}}{\sqrt{\pi\tau}})\nonumber\\
B^{III}_0&=&-\frac{\rme^{\rmi\frac{\pi}{4}}}{\sqrt{\pi\tau}}\frac{x-L}{2\tau}\rme^{\rmi\frac{(x-L)^2}{4\tau}}\nonumber\\
B^{III}_{-k}&=&-\rme^{\rmi\frac{(x-L)^2}{4\tau}}(k^2\rme^{y_{x-L,-k}^2}\;er\!f\!c(y_{x-L,-k})+\frac{\rme^{\rmi\frac{\pi}{4}}}{\sqrt{\pi\tau}}(-k+\frac{x-L}{2\tau}))\nonumber\\
B^{III}_n&=&-\rme^{\rmi\frac{(x-L)^2}{4\tau}}(p_n^2\rme^{y_{x-L,n}^2}\;er\!f\!c(y_{x-L,n})+
\frac{\rme^{\rmi\frac{\pi}{4}}}{\sqrt{\pi\tau}}(p_n+\frac{x-L}{2\tau}))\nonumber\\
S^{III}_k&=&-\rme^{\rmi\frac{(x-L)^2}{4\tau}}\rme^{y_{x-L,k}^2}\;er\!f\!c(y_{x-L,k})\nonumber\\
S^{III}_n&=&-\rme^{\rmi\frac{(x-L)^2}{4\tau}}\rme^{y_{x-L,n}^2}\;er\!f\!c(y_{x-L,n})\\
C^{III}_1&=&-\rme^{-\rmi\tau{k^2}}\;er\!f\!c(-\rmi\sqrt{\rmi\tau}k)\nonumber\\
C^{III}_2&=&-\rme^{-\rmi\tau{k^2}}\;er\!f\!c(\rmi\sqrt{\rmi\tau}k)\nonumber\\
C^{III}_3&=&-\rme^{\rmi\frac{(x-L)^2}{4\tau}}\rme^{y_{x-L,-k}^2}\;er\!f\!c(y_{x-L,-k})\nonumber\\
C^{III}_4&=&-\rme^{\rmi\frac{(x-L)^2}{4\tau}}\rme^{y_{x-L,k}^2}\;er\!f\!c(y_{x-L,k})\nonumber
\end{eqnarray}
In the equations (16), (18) and (20) the terms $[S]$ are the ones
arising in the shutter problem. The particular values of the Green
function involved are calculated in Appendix B.

\section{The short time limit}
The $t\rightarrow0$ limit is interesting both as a check of the
calculation above and for the study of the scattered wave at short
times. Here we aim only to recover the initial wave function
$\psi(x,0)= \rme^{\rmi kx}$ at $t=0$, which must happen in each of
the regions $I$, $II$ and $III$.

Careful use of the limiting values of $er\!f\!c(z)$ and/or
$w(z)=\rme^{-z^2}er\!f\!c(-\rmi z)$ for $z\rightarrow0$ and for
$z\rightarrow\infty$ in different directions of the complex $z$
plane must be made. Notice that $y_{x,q}$ tends to $\infty$ in
different directions for different $q$. Also the properties of the
set of resonant functions $u_n(x)$ as a basis of the space of
solutions are crucial in region II (Appendix G). Asymptotic
expressions for $er\!f\!c(z)$ and $w(z)$ can be used to obtain
approximations to the form of $\psi(x,t)$ in each region for small
values of $t$, a task that will be faced elsewhere.

In region $I$, the factor $A^I_n$ tends to $-2\rmi\delta(x)$ and
also $S^I_0\,$, $S^I_k$ and $S^I_n$ tend to a distribution
concentrated in $x=0\,$. The factors $B^I_{-k}\,$, $B^I_n\,$,
$S^I_3\,$ and $S^I_4\,$ vanish exactly in this limit and both
$S^I_1\,$ and $S^I_2\,$ tend to $-1\;$. Therefore, as required for
$x < 0\;$,
$$lim_{t\rightarrow0}\;\psi_{I}(x,t)= \rme^{\rmi kx}$$

In region $II$, the factors $A^{II}_n\,$, $B^{II}_{-k}\,$,
$B^{II}_n\,$, $S^{II}_k\,$ and $S^{II}_n$ tend to the value $-1$ so
that the $[A]$ term in (18) gives
\begin{equation}
\int_{0}^{L}\rmd x'\sum_{n}p_n\frac{u_n(x)u_n(x')}{N_n}\,\rme^{\rmi
kx'}=\rme^{\rmi kx}
\end{equation}
The $[B]$ terms yield
\begin{eqnarray}
\fl\rmi kG(L,x,-k)\rme^{\rmi kL}-&&\rmi\rme^{\rmi kL}\sum_{n}\frac{p_n}{k+p_n}\frac{u_n(L)u_n(x)}{N_n}\nonumber\\
&&=\rmi k\rme^{\rmi kL}\sum_{n}\frac{1}{-k-p_n}\frac{u_n(L)u_n(x)}{N_n}-\rmi\rme^{\rmi kL}\sum_{n}\frac{p_n}{k+p_n}\frac{u_n(L)u_n(x)}{N_n}\nonumber\\
&&=-\rmi\rme^{\rmi kL}\sum_{n}\frac{u_n(L)u_n(x)}{N_n}=0
\end{eqnarray}
and likewise for the $[S]$ terms:
\begin{eqnarray}
\fl\rmi kG(0,x,k)-\rmi \sum_{n}&&\frac{p_n}{k-p_n}\frac{u_n(0)u_n(x)}{N_n}\nonumber\\
&&=\rmi k\sum_{n}\frac{1}{k-p_n}\frac{u_n(0)u_n(x)}{N_n}-\rmi \sum_{n}\frac{p_n}{k-p_n}\frac{u_n(0)u_n(x)}{N_n}\nonumber\\
&&=\rmi \sum_{n}\frac{u_n(0)u_n(x)}{N_n}=0
\end{eqnarray}

In region $III$, the factor $A^{III}_n$ tends to $-2\rmi\delta(x-L)$
and also $B^{III}_0$, $B^{III}_{-k}$ and $B^{III}_n$ approach
distributions concentrated in $x=L\;$. Thus they vanish for $x>L$.
The factors $S^{III}$, $S^{III}_n$, $C^{III}_3$ and $C^{III}_4$
vanish exactly in this limit, whereas both $C^{III}_1$ and
$C^{III}_2$ tend to $-1$. Therefore the final result is that, for
$x>L\;$, also
$$lim_{t\rightarrow0}\;\psi_{III}(x,t)= \rme^{\rmi kx}$$

We have quoted singularities at the points $x=0$ and $x=L$
respectively in the regions $I$ and $III$, but these points must be
excluded from the domain of respectively $\psi_{I}(x,t)$ and
$\psi_{III}(x,t)\,$, as argued in Appendix F.

\section{The large time limit}

For $\tau\rightarrow\infty$ we see that also
$y_{x,q}\rightarrow\infty$ as before, but in yet different
directions of the complex plane for the different $q$.

In region I, the factors $A^I_n\,$, $B^I_{-k}\,$, $B^I_n\,$,
$S^I_0\,$, $S^I_n\,$, $S^I_2\,$ and $S^I_4\,$ vanish, whereas
\begin{eqnarray}
S^I_k&\rightarrow&-2 k^2\rme^{-\rmi E_kt}\rme^{-\rmi kx}\nonumber\\
S^I_1&\rightarrow&-2\rme^{-\rmi E_kt}\\
S^I_3&\rightarrow&-2\rme^{-\rmi E_kt}\rme^{-\rmi kx}\nonumber
\end{eqnarray}
Using equation (B.4), some of the $\rme^{-\rmi kx}$ terms cancel out
in (16) and one is left with
\begin{eqnarray}
lim_{t\rightarrow\infty}\;\psi_{I}(x,t)&=&\rme^{-\rmi
E_kt}(\rme^{\rmi
kx}+R(k)\rme^{-\rmi kx})\\
&=&\rme^{-\rmi E_kt}\phi^{in}_r(x)\nonumber
\end{eqnarray}
that is, the asymptotic scattering {\it in} solution, where
$E_k=k^2/\,2m\;$ (see Appendix H).

In region $II$, the factors $A^{II}_n$, $B^{II}_{-k}$, $B^{II}_n$
and $S^{II}_n$ vanish, whereas $S^{II}_k\rightarrow
-2\rme^{-\rmi\tau k^2}$. Therefore, using (B.2), for $0\leq x\leq L$
we have
\begin{eqnarray}
lim_{t\rightarrow\infty}\;\psi_{II}(x,t)&=&2\rmi k\;G(0,x,k)\rme^{-\rmi\tau k^2}\nonumber\\
&=&\rme^{-\rmi E_kt}\phi^{in}_r(x)
\end{eqnarray}
Thus the infinite plane wave evolves into the same final state of
the shutter initial condition also in region $II$.

For the region $III$, the only non-vanishing factors are
\begin{eqnarray}
S^{III}_k&\rightarrow&-2\;\rme^{\rmi k(x-L)}\rme^{-\rmi\tau k^2}\nonumber\\
C^{III}_2&\rightarrow&-2\;\rme^{-\rmi\tau k^2}\\
C^{III}_4&\rightarrow&-2\;\rme^{\rmi k(x-L)}\rme^{-\rmi\tau
k^2}\nonumber
\end{eqnarray}
Then the terms corresponding to $C^{III}_2$ and $C^{III}_4$ in (20)
cancel each other and, again using (B.2), one is left with the same
asymptotic solution of the shutter
\begin{eqnarray}
lim_{t\rightarrow\infty}\;\psi_{III}(x,t)&=&2\rmi k\;G(0,L,k)\rme^{\rmi k(x-L)}\rme^{-\rmi\tau k^2}\nonumber\\
&=&T(k)\rme^{\rmi kx}\rme^{-\rmi E_kt}\\
&=&\rme^{-\rmi E_kt}\phi^{in}_r(x)\nonumber
\end{eqnarray}

The finite time behaviour is thus made up of transient modes which
quickly dampen out, leaving (for $k>0$) the scattering asymptotic
solution with outgoing BC at $x=L$.

\section{Conclusions}

A solution for the time evolution of an infinite plane wave in the
presence of a simple square barrier has been worked out for each of
the regions of the potential and Section 4 is the main result of
this work. Among other terms, this solution contains a sum of
explicit analytical contributions corresponding to each of the
(infinitely many) resonance poles. As in previous works in related
problems \cite{Muga1} \cite{Calderon1}, only the location of these
poles needs to be obtained by numerical methods.

In the solution obtained the shutter terms have been pinpointed.
Similarly, the contributions to the time-evolved wave function
coming from the segments of the initial wave function lying inside
and at the right of the barrier can be identified.

As the main application of this knowledge we envisage the
possibility of studying the enhancement or the suppression of the
transmission of the single Fourier components of any realistic wave
packet. This should provide new detailed insight on interesting
phenomena like the super-luminal tunnelling \cite{Leon}, the
breakdown of energy conservation \cite{Muga2} by transient
interference in wave packet collisions with barriers \cite{Muga3} or
the rising of forerunners.

The explicit solution obtained for finite time is useful for
deriving approximations valid for short times, hence for the study
of transient structures and forerunners. An immediate result of the
work is that the large time limit yields the same stationary
solution of the shutter. This again shows that the resonances
contribute only to transient structures of the scattered wave.

\ack{Work supported by MEC projects BFM2002-00834 and FIS2005-05304.
The authors are indebted to J. Le\'on for suggestions and helpful
discussions. J. Julve acknowledges the hospitality of the
Dipartimento di Fisica dell'Universit\`a di Bologna, Italy, where
part of this work was done.}

\appendix

\section{Resonant solutions}
We adopt units such that $\hbar=1\;$. The homogeneous equation
$\,L_xu(x)=0$ with the RBCs (3) has solutions only for a denumerable
set of values $p_n\,$ lying in the lower complex plane $p\,$ and
satisfying the condition
\begin{equation}
D(p)\equiv\ominus^2\rme^{\rmi p'L}-\oplus^2\rme^{-\rmi p'L}=0\;,
\end{equation}
where $p'\equiv\sqrt{p^2-2mV}\;$, $\oplus\equiv p+p'\;$ and
$\ominus\equiv p-p'\;$.  One can check that if $p_n$ is a solution,
then $-p_n^*$ is too, so that these values are in symmetrical
locations with respect to the imaginary axis. We let the label $n$
take integer values ($n\neq0$), with the growing positive $n$
indicating the $p_n$ with growing real positive part and
$p_{-n}\equiv -p_n^*$. One finds that $-\frac{\pi}{4}<ar\!g\,p_n<0$
and $\pi<ar\!g\,p_{-n}<\frac{5\pi}{4}$.

The solutions of $L_xu(x)=0$ are
\begin{eqnarray}
u_n(x)=&&\theta(-x)(-2p'_n)\rme^{-\rmi p_nx}\nonumber\\
&+&\theta(x)\theta(L-x)(\ominus_n\,\rme^{\rmi p'_nx}-\oplus_n\,\rme^{-\rmi p'_nx})\\
&+&\theta(x-L)\,(\ominus_n\,\rme^{-\rmi\ominus_nL}-\oplus_n\,\rme^{-\rmi\oplus_nL})\,\rme^{\rmi
p_nx}\;,\nonumber
\end{eqnarray}
up to an arbitrary multiplicative function of $p_n\;$.

The residues $C_n(x,x')=u_n(x)u_n(x')/N_n$ in (11) correspond to the
following choice of (complex) "norm" \cite{Calderon2}
\begin{equation} N_n=\rmi(u^2_n(0)+u^2_n(L))+2p_n\int^L_0\rmd
x\;u^2_n(x)
\end{equation}
which takes the value $N_n=-8mV(p_n\,L+2\rmi)$ for the solutions
(A.2).

\section{Analytical Green function}

The analytical solution to the Green equation
$\;L_xG(x,y,p)=\delta(x-y)\;$ for $x$ and $y$ in the interval
$[0,L]\,$, and obeying the RBCs (7), can be directly constructed:

\begin{eqnarray}
\fl G(x,y,p)=\frac{\rmi}{2p'}\frac{1}{D(p)}\;&\{&\;2mV\,(\rme^{\rmi p'(L-(x+y))}+\rme^{-\rmi p'(L-(x+y))})\\
&&\!\!\!\!-\oplus^2\rme^{-\rmi p'L}(\theta(y-x)\rme^{\rmi p'(y-x)}+\theta(x-y)\rme^{\rmi p'(x-y)})\nonumber\\
&&\!\!\!\!-\ominus^2\rme^{\rmi p'L}(\theta(y-x)\rme^{-\rmi
p'(y-x)}+\theta(x-y)\rme^{-\rmi p'(x-y)})\,\}\;,\nonumber\\\nonumber
\end{eqnarray}
where the symmetry $x\leftrightarrow y$ is explicit.

The limit $|p|\rightarrow\infty$ in different directions of the
complex plane can be directly read out in (B.1). For any values of
$x$ and $y$ in the interval [$0,L$] the Green function vanishes as
$\;1/p\;$ or faster in any direction, with the only exception of
$G(0,0,p)$ and $G(L,L,p)\,$, which grow as $p$ in the lower half
plane, while still decreasing as $\;1/p\;$ in the real axis.

The particular cases $G(0,x,p)$ and $G(L,x,p)$ are related to the
scattering solutions $\phi^{in}(x)$ (Appendix H). From the integral
formula $\;\int^L_0\rmd
x\,[\phi\,(L_xG-\delta(x-x'))-G\,L_x\phi]=0\;$ and the use of the
BCs for $\phi$ and the RBCs for $G\,$, one obtains \cite{Calderon4}
\begin{eqnarray}
G(0,x,p)&=&\frac{-\rmi}{2p}\;\sqrt{2\pi}\sqrt{\frac{p}{m}}\;\phi^{in}_r(x)\hskip 1cm (0<x\leq L)\\
G(L,x,p)&=&\frac{-\rmi}{2p}\;\sqrt{2\pi}\sqrt{\frac{p}{m}}\;\phi^{in}_l(x)\hskip
1cm (0\leq x<L)\\\nonumber
\end{eqnarray}
If $\rm{Im}\,p \geq 0\;$, the limits $x=0$ in (B.2) and $x=L$ in
(B.3) can be taken, namely
\begin{equation}
G(0,0,p)=\phi^{in}_r(0)=\frac{-\rmi}{2p}\;\sqrt{2\pi}\;(1+R(p))=\phi^{in}_l(L)=G(L,L,p)
\end{equation}

\section{Pole expansion and substractions}
Whenever the Green function $G(x,x',p)$ vanishes as $\;1/p\;$ in the
limit $|p|\rightarrow\infty$, the Mittag-Leffler theorem applies and
the validity of the pole expansion (11) is assured. This is not the
case for $G(0,0,p)$ and $G(L,L,p)\,$, and a substraction technique
must be used \cite{Nussenzveig}. This is performed by using the
contour integral

\begin{equation}
0=\frac{1}{2\pi \rmi}\oint_{\Gamma}\rmd
z\,\frac{1}{z^2}\frac{G(x,x',z)}{z-p}
\end{equation}
where the contour $\Gamma\equiv\{C_o,C_p\,,C_n,C_s\}$ consists of
small closed paths counterclock-wise encircling the poles of the
integrand (namely $C_o$ around the double pole at $z=0$, $C_p$
around $p$ and $C_n$ around each $p_n$) plus a large circle $C_s$
clock-wise encircling all these poles. The factor $z^{-2}$ has been
introduced to assure that the integrand goes as $z^{-2}$ even for
$G(0,0,p)$ and $G(L,L,p)\,$ so that the integral over $C_s$ vanishes
when the radius of the circle grows to $\infty$.

From (C.1) one readily obtains (see also \cite{Calderon5})
\begin{equation}
\fl
G(x,x',p)=p^2\sum^{\infty}_n\frac{1}{N_n}\frac{1}{p^2_n}\frac{u_n(x)u_n(x')}{p-p_n}
+[G(x,x',0)+p\;\partial_pG(x,x',p)\!\mid_{p=0}]
\end{equation}
When the pole expansion (11) holds, (C.2) becomes a trivial
identity, whereas for $G(0,0,p)$ and $G(L,L,p)\,$ a holomorphic
(linearly growing with $|p|$ ) part remains.

\section{Inverse Laplace transform}

The inverse Laplace transform in the variable $s$ in (12), involves
an integration in the complex $s$-plane along a line parallel to the
imaginary axis with $\rm{Re}\,c>0$. In the $p$-plane (recall
$p=\sqrt{\rmi}\sqrt{2m\,s}$), this path translates to a
hyperbola-like one with asymptotes in the positive real and
imaginary axis. For $t>0\,$, the factor
$\rme^{-\rmi\frac{p^2}{2m}t}$ assures that it can be deformed into
an integration from $+\infty$ to $-\infty$ along (and slightly
above) the real axis if the integrand has poles only in the lower
half plane (and on the real axis). For $t<0\,$, the path can be
closed along a quarter circle of large radius in the 1st quadrant,
thus enclosing a region without poles and giving $\psi(x,t)=0$,
consistently with the Laplace method and causality
\cite{Nussenzveig}.

\section{Pole expansion of $p\;\bar\psi(x,p)$}
We rewrite (8) as
\begin{equation}
\fl\bar\psi_{II}(x,p)\;=\;\rmi\alpha\int_{0}^{L}\rmd
x'\,f^{II}_1(x',x,p)\,\rme^{\rmi kx'} -\alpha\;\rme^{\rmi
kL}f^{II}_2(x,p)-\alpha\, f^{II}_3(x,p)\\\nonumber
\end{equation}
where
$$\;f^{II}_1(x',x,p)=G(x',x,p)\;,
\;f^{II}_2(x,p)=\frac{G(L,x,p)}{p+k}\; {\rm and}
\;f^{II}_3(x,p)=\frac{G(0,x,p)}{p-k}\;.$$ According to the
asymptotic behaviour of the $\;f^{II}_i(p)\;$ ($i=1,2,3$) above we
consider the contour integrals
\begin{equation}
0=\frac{1}{2\pi \rmi}\oint_{\Gamma_i}\rmd
z\,\frac{z^{n_i}}{z-p}f^{II}_i(z)\nonumber
\end{equation}
where the values $n_1=0$, and $n_2=n_3=1\;$ are assigned so that the
integrand of (E.2) behaves as $z^{-2}$ for large $z\;$. The contours
$\Gamma_i$ include circles around the poles in the respective
integrand (namely $p\,$, $p_n$, $k$ or $-k\;$) plus the large circle
$C_s\,$, as in Appendix C. For $i=1$, equation (E.2) yields
$f^{II}_1(p)$, which must be multiplied by $p$ later on, whereas for
$i=2,3$ one directly obtains $p\,f^{II}_i(p)$. Collecting these
results and using the pole expansion of the Green function, (14) is
obtained.

The derivation of (13) and (15) follows the same lines. There we
define
$$\;f^{I}_1(p)=G(x',0,p)\;,
\;f^{I}_2(p)=\frac{G(0,0,p)}{p-k}\;,
\;f^{I}_3(p)=\frac{G(L,0,p)}{p+k}\;,$$
and
$$\;f^{III}_1(p)=G(x',L,p)\;,
\;f^{III}_2(p)=\frac{G(L,L,p)}{p+k}\;,
\;f^{III}_3(p)=\frac{G(0,L,p)}{p-k}\;,$$ where the powers required
re $n_1=0$, $n_2=-1$ and $n_3=1\;$. Notice that the negative power
$n_2=-1$, needed by the asymptotic behaviour of $G(0,0,z)$ and
$G(L,L,z)\,$, introduces an extra pole at $z=0$ and is related to
the discussion of Appendix C.

\section{{\it erfc(z)}-related integrals}
In the computation of $\psi_{I}(x,t)$,$\psi_{II}(x,t)$ and
$\psi_{III}(x,t)$, the following types of integrals arise:
\begin{eqnarray}
I_0(x)&=&\frac{1}{\rmi\pi}\int^{+\infty}_{-\infty}\rmd p\;\rme^{-\rmi\tau p^2}\rme^{\rmi xp}=\frac{\rme^{-\rmi\frac{3\pi}{4}}}{\sqrt{\pi\tau}}\rme^{\rmi\frac{x^2}{4\tau}}\\
I_1(x)&=&\frac{1}{\rmi\pi}\int^{+\infty}_{-\infty}\rmd p\;p\;\rme^{-\rmi\tau p^2}\rme^{\rmi xp}=\frac{\rme^{-\rmi\frac{3\pi}{4}}}{\sqrt{\pi\tau}}\frac{x}{2\tau}\rme^{\rmi\frac{x^2}{4\tau}}\\
I_0(x,q)&=&\frac{1}{\rmi\pi}\int^{+\infty}_{-\infty}\rmd p\;\frac{1}{p-q}\,\rme^{-\rmi\tau p^2}\rme^{\rmi xp}\equiv -2\,M(x,q,t)=-\rme^{\rmi\frac{x^2}{4\tau}}\rme^{y^2_{x,q}}\;er\!f\!c(y_{x,q})\ \ \ \ \ \ \ \\
I_1(x,q)&=&\frac{1}{\rmi\pi}\int^{+\infty}_{-\infty}\rmd p\;\frac{p}{p-q}\,\rme^{-\rmi\tau p^2}\rme^{\rmi xp}=I_0(x)+qI_0(x,q)\\
I_2(x,q)&=&\frac{1}{\rmi\pi}\int^{+\infty}_{-\infty}\rmd
p\;\frac{p^2}{p-q}\,\rme^{-\rmi\tau p^2}\rme^{\rmi x
p}=I_1(x)+qI_0(x)+q^2I_0(x,q)\;,
\end{eqnarray}
where
$$M(x,q,t)\equiv\frac{-1}{2\pi
\rmi}\int^{+\infty}_{-\infty}\rmd
p\;\frac{1}{p-q}\,\rme^{-\rmi\frac{t}{2m} p^2}\rme^{ixp}$$ is the
Moshinsky function, often found in the literature.

By means of the change of variables $p\rightarrow p-\frac{x}{2\tau}$
they can be related to the integral representation of the $error\,
f\!unction$ \cite{Stegun}
\begin{equation} w(z)\equiv \rme^{-z^2} er\!f\!c(-\rmi z)=
-\frac{\rmi}{\pi}\int^{+\infty}_{-\infty}du\;\frac{\rme^{-u^2}}{u-z}
\hskip 2cm \rm{Im}\,z>0
\end{equation}
so that
$$M(x,q,t)=\frac{1}{2}\rme^{\rmi\frac{m}{2t}x^2}w(\rmi\rme^{-\rmi\frac{\pi}{4}}\sqrt{\frac{m}{2t}}(x-\frac{t}{m}
q))\;.$$ In the expressions above, is convenient the use of the
variables
$\;y_{x,q}\equiv\frac{\rme^{-\rmi\frac{\pi}{4}}}{\sqrt{4\tau}}(x-2\tau
q)\,$, where $\tau\equiv\frac{t}{2m}$

When converting the exponent $\rmi\tau p^2$ into $u^2$, namely
$u=\sqrt{\rmi\tau}p\,$, an integration along a line running between
$\pm \sqrt{\rmi}\,\infty$ results. In the case $q=p_n\,$, rotating
this path by an angle $\pi/4$ towards the real axis crosses the pole
$\sqrt{\rmi\tau}\,p_n$, contributing a residue. For $q=\pm k$, a
small negative imaginary part is supposed (see Appendix D). The case
$\rm{Im}\,z<0$ can be tackled by using the property
$w(z^*)=w^*(-z)$.

The results of the integrals above, as given in (17), (19) and (21),
are
\begin{equation}
\fl \hskip 1.0cm
\begin{array}{rcl}
A^I_n&=&I_1(-x,p_n)\\
B^I_{-k}=S^I_4&=&I_0(-x,-k)\\
B^I_n=S^I_n&=&I_0(-x,p_n)\\
S^I_0&=&I_1(-x)\\
S^I_k&=&I_2(-x,k)\\
S^I_3&=&I_0(-x,k)\\
\end{array}
\hskip 1.0 cm
\begin{array}{rcl}
A^{II}_n&=&I_1(0,p_n)\nonumber\\
A^{III}_n&=&I_1(x-L,p_n)\nonumber\\
B^{III}_0&=&I_1(x-L)\nonumber\\
B^{III}_{-k}&=&I_2(x-L,-k)\nonumber\\
B^{III}_n&=&I_2(x-L,p_n)\nonumber\\
S^{III}_k&=&I_0(x-L,k)\\
S^{III}_n&=&I_0(x-L,p_n)\nonumber\\
S^I_2=B^{II}_{-k}=C^{III}_1&=&I_0(0,-k)\nonumber\\
S^I_1=S^{II}_k=C^{III}_2&=&I_0(0,k)\nonumber\\
C^{III}_3&=&I_0(x-L,-k)\nonumber\\
C^{III}_4&=&I_0(x-L,k)\nonumber\\
B^{II}_n=\;S^{II}_n&=&I_0(0,p_n)\nonumber
\end{array}
\end{equation}
The computation needs some care in the limit $\tau\rightarrow 0$
when we simultaneously let $x=0\,$, as is the case for
$A^{II}_n=I_1(0,p_n)$, because the change of variables $p\rightarrow
p-\frac{x}{2\tau}$ becomes ill defined. In fact, taking first
$x=0\,$, the naive result $A^{II}_n=-p_n\rme^{-\rmi \tau
p^2_n}\;er\!f\!c(\rmi\sqrt{\rmi
\tau})-\frac{\rme^{\rmi\frac{\pi}{4}}}{\sqrt{\pi\tau}}$ is singular
when $\tau\rightarrow 0\;$. The correct result is attainable as
follows: The troublesome first term in (18) describes the time
evolution in region II of the initial condition
$\psi_0(x)=\rme^{\rmi kx}$ governed by the retarded propagator
$G^R(x,x',t)$ which is the inverse Laplace transformed of the
$G(x,x',p)$ (subject to RBCs). As such, it has the expansion
$$G^R(x,x',t)=\sum_{n}\rme^{-\rmi\frac{p^2_n}{2m}t}\,p_n\frac{u_n(x)u_n(x')}{N_n}\;,$$
hence the result for $A^{II}_n$ quoted in (19). Performing the
inverse Laplace transform by integration in the variable $p$ is not
convenient in this case, whereas it is more direct in the variable
$s=-\rmi\frac{p^2}{2m}=-\rmi E_p\;$.

The singularities for $\tau\rightarrow 0$ reported in Section 5 have
the same origin and, in this sense, the points $x=0$ and $x=L$ must
be considered as belonging to the non trivial region II.

\section{$t\rightarrow 0,\infty$ limits of $\rme^{y^2_{x,q}}\,er\!f\!c(y_{x,q})$}

We see that, for $t\rightarrow 0\,$,
$$y_{x,q}\sim \rme^{-\rmi\frac{\pi}{4}}\sqrt{\frac{m}{2t}}\;\;x\;\rightarrow \rme^{-\rmi\frac{\pi}{4}}\cdot\infty\;,$$
regardless of $q$ being equal to $\pm k$, $p_n\,$ or $p_{-n}\,$.

For $t\rightarrow \infty\,$ one has
$$y_q\sim \rme^{\rmi\,\frac{3\pi}{4}}\sqrt{\frac{t}{2m}}\;\;q\;\rightarrow\; \rme^{\rmi\phi_q}\cdot\infty\;,$$
where $\;\frac{\pi}{2}<\phi_n<\frac{3\pi}{4}\;\;$,
$\;-\frac{\pi}{4}<\phi_{-n}<0\;\;$,  $\phi_k=\frac{3\pi}{4}\;\;$ and
$\;\phi_{-k}=-\frac{\pi}{4}\;\;$.

The following asymptotic formulas \cite{Stegun} apply:
$$lim_{z\rightarrow\infty}\;er\!f\!c(z)=0\hskip 2cm|ar\!g\,z|<\frac{\pi}{4}$$
$$lim_{z\rightarrow\infty}\;\rme^{z^2}er\!f\!c(z)\sim\frac{1}{\sqrt{\pi}}\frac{1}{z}(1+\sum_{m=1}^{\infty}\frac{C_m}{z^{2m}})\rightarrow 0\hskip 2cm|ar\!g\,z|<\frac{3\pi}{4}$$
Only the case $\phi_k=\frac{3\pi}{4}\;$ is out of their range, in
which case the relationship
$$\rme^{y^2_{x,k}}\,er\!f\!c(y_{x,k})=2\rme^{y^2_{x,k}}-\rme^{y^2_{x,k}}\,er\!f\!c(-y_{x,k})\;,$$
where $|ar\!g(-y_{x,k})|=\frac{\pi}{4}\,$, is helpful.

\section{Scattering solutions}

The usual scattering solutions of $L_x\bar\psi(x,p)=0$ for
continuous real energies $E_p=p^2/\,2m\,>0\;$ may be characterized
according to the homogeneous BC they obey. Contrary to the resonant
solutions, which are subject to {\it two} BCs,  each one of the
scattering solutions solution is subject to {\it one} homogeneous
BC, alternatively at $x=0$ or at $x=L$ :
\begin{eqnarray}
\partial_x\phi^{in}_r(x)\!|_L&=&\rmi p\,\phi^{in}_r(L)\nonumber\\
\partial_x\phi^{in}_l(x)\!|_0&=&-\rmi p\,\phi^{in}_l(0)\nonumber\\
\partial_x\phi^{out}_r(x)\!|_0&=&\rmi p\,\phi^{in}_r(0)\nonumber\\
\partial_x\phi^{out}_l(x)\!|_L&=&-\rmi p\,\phi^{in}_l(L)\;,\\\nonumber
\end{eqnarray}
where $r$($l$) labels the right(left)-moving character of the
impinging wave. Notice that the reverse notation is often found in
the literature, with  $r$($l$) indicating the wave coming from the
right(left) of the barrier.

These solutions, for instance
$$\phi^{in}_r(x,p)=\frac{1}{\sqrt{2\pi}}\sqrt{\frac{m}{p}}\;[\;\theta(-x)(\rme^{\rmi px}+R\,\rme^{-\rmi px})\nonumber\\
+\theta(x)\theta(L-x)(P\,\rme^{\rmi p'x}+Q\,\rme^{-\rmi p'x})
+\theta(x-L)\,T\,\rme^{\rmi px}\,]\,,$$  are $\delta$-normalized in
energy.

\end{document}